\documentclass[sigconf]{acmart-me}

\usepackage{booktabs} 
\usepackage{url}
\usepackage{color}
\usepackage{enumitem}
\setcopyright{rightsretained}

\acmDOI{}

\acmISBN{}

\acmConference[MediaEval'21]{Multimedia Evaluation Workshop}{December 13-15 2021}{Online} 
\acmYear{2021}
\copyrightyear{}

\acmPrice{}

\begin{document}
\title{Overview of the EEG Pilot Subtask at MediaEval 2021: Predicting Media Memorability}

\author{Lorin Sweeney\textsuperscript{1}, 
Ana Matran-Fernandez\textsuperscript{2}, 
Sebastian Halder\textsuperscript{2},\\ 
Alba G. Seco de Herrera\textsuperscript{2},
Alan Smeaton\textsuperscript{1}, 
Graham Healy\textsuperscript{1}}
\affiliation{\textsuperscript{1}School of Computing, Dublin City University\\ \textsuperscript{2}School of Computer Science and Electronic Engineering, 
University of Essex}
\email{lorin.sweeney8@mail.dcu.ie, {amatra, s.halder, alba.garcia}@essex.ac.uk, {alan.smeaton, graham.healy}@dcu.ie}

\renewcommand{\shortauthors}{Sweeney et al.}
\renewcommand{\shorttitle}{Predicting Media Memorability}

\begin{abstract}
The aim of the Memorability-EEG pilot subtask at MediaEval'2021 is to promote interest in the use of neural signals---either alone or in combination with other data sources---in the context of predicting video memorability by highlighting the utility of EEG data. The dataset created consists of pre-extracted features from EEG recordings of subjects while watching a subset of videos from Predicting Media Memorability subtask 1.
This demonstration pilot gives interested researchers a sense of how neural signals can be used without any prior domain knowledge, and enables them to do so in a future memorability task. The dataset can be used to support the exploration of novel machine learning and processing strategies for predicting video memorability, while potentially increasing interdisciplinary interest in the subject of memorability, and opening the door to new combined EEG-computer vision approaches.

\end{abstract}

%
%
%
%
%

\maketitle

\section{Introduction and related work}
\label{sec:intro}
Even though the nature and constitution of people’s memories remains elusive, and our understanding of what makes one thing more/less memorable than another is still nascent, combining computational (e.g., machine learning) and neurophysiological (e.g., electroencephalography; EEG) tools to investigate the mechanisms (formation and recall) of memory may offer insights that would be otherwise unobtainable. While EEG is not a tool that can directly explain what makes a video more/less memorable, it can help us trim the umbral undergrowth surrounding the subject, shedding light and offering a potential leap forward in our understanding of the interplay between the mechanisms of memory and memorability. 

The purpose of this pilot study at MediaEval'2021 \cite{ME2021} was to collect enough EEG data for proof of concept and demonstration purposes, showcasing what could be done in subsequent work on predicting media memorability. The study involved the collection, filtering, and interpretation of neurophysiological data, and the use and evaluation of machine learning methods to enable the assessment of EEG data as a predictor of video memorability. 
The study has culminated in a demonstration of the utility of EEG in the context of video memorability, along with the public release of processed EEG features for others to explore\footnote{Dataset and examples of use, as well as the code to replicate the results in this paper, are available at \url{https://osf.io/zt6n9/}}. This study has the potential to not only broaden the research horizons of computing researchers, allowing them to explore and leverage EEG features without any of the requisite domain knowledge, but also increase the interdisciplinary interest in the subject of memorability more broadly.

Applying EEG to the question of whether an experience will be subsequently remembered or forgotten is a well researched area \cite{sanquist1980electrocortical, karis1984p300, klimesch1999eeg, noh2014using}. Memorability, however, has been shown to be distinct from subsequent memory effects \cite{rugg2007event, bainbridge2017memory}, and received little interdisciplinary attention. Additionally, even though the application of machine learning to EEG is an active area of interest---allowing for the automation or augmentation of neurological diagnostics \cite{ieracitano2020novel, lehmann2007application, hosseinifard2013classifying, engemann2018robust}, and the classification of emotional states \cite{wang2014emotional}, mental tasks \cite{liang2006classification}, and sleep stages \cite{ebrahimi2008automatic}---the use of EEG to predict visual memorability has yet to be firmly established, and was previously limited to static content \cite{jo2020prediction}. To the best of our knowledge, this paper outlines the first application of EEG to video memorability.


\section{Experiment design and structure}
The stimuli used in the study are a subset of the subtask 1 data (i.e., the short-term video memorability prediction task) in MediaEval'2021 \cite{ME2021}, and consists of 450 videos, 96 of which were designated as targets and selected to reflect the bottom and top 50 memorable videos from the TRECVid dataset, 200 were selected to reflect the next top and bottom 100, and 100 were selected to reflect the middle 100 memorable videos (95 selected + 5 duplicates) from the set of subtask videos. EEG data was collected from 11 subjects while they completed a short-term memory experiment, which was used to annotate the videos for memorability. EEG data acquisition\footnote{Data collection for participants 1--5 was carried out at Dublin City University (DCU) with approval from the university's Research Ethics Committee (DCUREC / 2021 / 171), and for participants 6--11 at the University of Essex (UoE) with approval from the Ethics Committee (ETH2122-0001). Data at DCU was collected using a 32-channel ANT Neuro eego system with a sampling rate of 1000~Hz. Data at UoE was collected using a 64-channel BioSemi ActiveTwo system at a sampling rate of 2048~Hz.} was carried out in two separate locations using a shared experimental procedure, and each location annotated the same set of videos. 
Rather than being split into separate encoding and recognition phases, the experiment was continuous in nature.

Before the experiment was carried out, participants were given a verbal description of the experiment procedure, presented with a set of written instructions, and taken through a practice run of 3 videos to familiarise them with the experiment.
The experiment used a total of 450 videos, 192 of which were the target videos (96 targets, shown twice), and the remaining 258 videos were the fillers. The experiment was broken into 9 blocks of 50 videos, where a fixation cross was displayed for 3--4.5s, followed by the video presentation for its \textasciitilde 6 second duration, followed by a ``get ready to answer'' prompt of 1--3 seconds, followed by a 3s period for recognition response (repeated video or not). The time per block was approximately 700 seconds (\textasciitilde 12 minutes) without accounting for 30-second closed/open eye baselines and breaks, which occurred between blocks. In order to account for recency effects, the first 50 videos presented did not include targets, but had 5 filler repeats, and the presentation positions of targets between each of the participants was pseudo-randomised, with the distances between target and repeat videos roughly fitting a uniform distribution, and the position of each block aside from block 1 being rotated by 1 for each participant. 

\section{Analysis and Results}

EEG data from both locations were processed in the same way for the 30~channels that were common across the two setups: data were first referenced using a common average and band-pass filtered between 0.1--30~Hz using a symmetric linear-phase FIR filter. Independent Component Analysis (ICA) was used to remove artifacts, and trial rejection using subject-specific thresholds was applied.

To establish a baseline using features extracted from the time domain, the EEG was low-pass filtered with a cutoff frequency of 15~Hz and downsampled to 30~Hz. We applied baseline correction to the average of the 250-ms  pre-stimulus interval and extracted the data corresponding to the first second of each repeated clip, from each of the 30~channels, and concatenated it to form a feature vector. We term these the Event-Related Potential (ERP) features.

A second set of features were extracted from the EEG, this time from the time-frequency domain, which we refer to as ERSP (Event-Related Spectral Perturbation) features. For this, we extracted 4-second long epochs and computed a trial-by-trial time-frequency representation using Morlet wavelets for frequencies between 2-30~Hz. For this set of features, we used data from only 4~channels, namely Fz, Cz, Pz, and O1.


Since there were very few forgotten clips, in this task we differentiate between the first and the second viewing of clips that were successfully remembered based only on EEG data.
To establish a baseline, we standardised the data to have mean zero and unit standard deviation, and used \textit{scikit-learn}'s Bayesian Ridge regressor with default parameters. Results were obtained through 20-fold cross-validation with a 20\% train-test split, separately for ERP and ERSP features. The individual classification results for each participant are shown in Table~\ref{table:aucs}, measured using Area Under the Receiver Operating Characteristic Curve \cite{pedregosa2011scikit}.

\begin{table}
\caption{Mean AUC values obtained for each participant across all folds, separately for ERP and ERSP features.}
\label{table:aucs}
\centering
\begin{tabular}{c c c}
\toprule
Participant & ERP-based classification & ERSP-based classification\\
\midrule
1 & 0.564 $\pm$ 0.09 & 0.522 $\pm$ 0.09 \\
2 & 0.585 $\pm$ 0.11 & 0.558 $\pm$ 0.07 \\
3 & 0.520 $\pm$ 0.07 & 0.532 $\pm$ 0.07 \\
4 & 0.666 $\pm$ 0.07 & 0.626 $\pm$ 0.09 \\
5 & 0.714 $\pm$ 0.06 & 0.649 $\pm$ 0.08 \\
6 & 0.555 $\pm$ 0.11 & 0.522 $\pm$ 0.10 \\
7 & 0.601 $\pm$ 0.10 & 0.525 $\pm$ 0.08 \\
8 & 0.590 $\pm$ 0.08 & 0.674 $\pm$ 0.08 \\
9 & 0.609 $\pm$ 0.09 & 0.489 $\pm$ 0.06 \\
10 & 0.628 $\pm$ 0.06 & 0.618 $\pm$ 0.09 \\
11 & 0.477 $\pm$ 0.08 & 0.611 $\pm$ 0.12 \\
\midrule
Mean & 0.591 $\pm$ 0.06 & 0.575 $\pm$ 0.06 \\
\bottomrule
\end{tabular}
\end{table}

\begin{figure}
    \centering
    \includegraphics[clip,trim=0 0 0 0, width=0.5\textwidth]{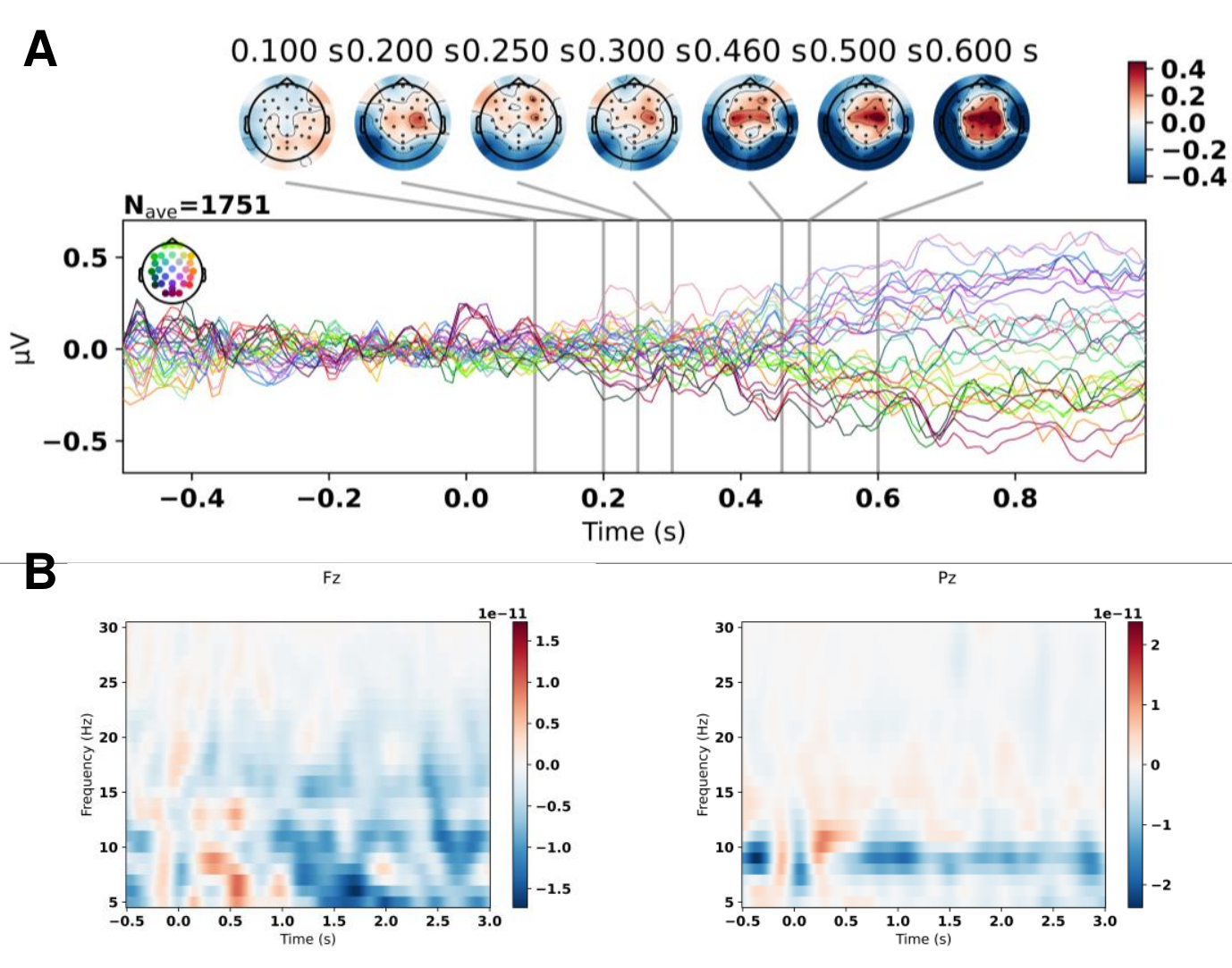}
    \caption{Grand-averaged butterfly plot showing  differences in EEG activity for the second  minus  first presentation for videos for the first second (top-A). Averaged time-frequency differences in power for the second presentation minus that for the first presentation of videos for the first 3 seconds for channels Fz and Pz (bottom-B left and right, resp.). }
    \label{fig:my_label}
\end{figure}




%

\section{Discussion and Outlook}
This was an exploratory pilot task to guide the development of a future experimental protocol for capturing EEG signatures relating to successful memory encoding and retrieval to be used in predicting video memorability. While our experimental protocol resulted in too little data to examine differences between successful and unsuccessful encoding, we show EEG-related differences exist between the encoding and recognition phases of previously seen videos. These results indicate that EEG signatures relating to memory processes for video are present, and thus suitable to be collected with a revised experimental protocol and more participants to support a future fully-fledged task for predicting video memorability. The preprocessed EEG data captured is released to the research community.

\begin{acks}
This work was part-funded by NIST Award No. 60NANB19D155 and by Science Foundation Ireland under grant number SFI/12/RC/2289{\_P2}.
\end{acks}

\bibliographystyle{ACM-Reference-Format}
\def\bibfont{\small} 
\bibliography{sigproc} 


\begin{thebibliography}{00}


\ifx \showCODEN    \undefined \def \showCODEN     #1{\unskip}     \fi
\ifx \showDOI      \undefined \def \showDOI       #1{#1}\fi
\ifx \showISBNx    \undefined \def \showISBNx     #1{\unskip}     \fi
\ifx \showISBNxiii \undefined \def \showISBNxiii  #1{\unskip}     \fi
\ifx \showISSN     \undefined \def \showISSN      #1{\unskip}     \fi
\ifx \showLCCN     \undefined \def \showLCCN      #1{\unskip}     \fi
\ifx \shownote     \undefined \def \shownote      #1{#1}          \fi
\ifx \showarticletitle \undefined \def \showarticletitle #1{#1}   \fi
\ifx \showURL      \undefined \def \showURL       {\relax}        \fi
\providecommand\bibfield[2]{#2}
\providecommand\bibinfo[2]{#2}
\providecommand\natexlab[1]{#1}
\providecommand\showeprint[2][]{arXiv:#2}

\bibitem[\protect\citeauthoryear{Bainbridge, Dilks, and Oliva}{Bainbridge
  et~al\mbox{.}}{2017}]%
        {bainbridge2017memory}
\bibfield{author}{\bibinfo{person}{Wilma~A Bainbridge},
  \bibinfo{person}{Daniel~D Dilks}, {and} \bibinfo{person}{Aude Oliva}.}
  \bibinfo{year}{2017}\natexlab{}.
\newblock \showarticletitle{Memorability: A stimulus-driven perceptual neural
  signature distinctive from memory}.
\newblock \bibinfo{journal}{{\em NeuroImage\/}}  \bibinfo{volume}{149}
  (\bibinfo{year}{2017}), \bibinfo{pages}{141--152}.
\newblock


\bibitem[\protect\citeauthoryear{Ebrahimi, Mikaeili, Estrada, and
  Nazeran}{Ebrahimi et~al\mbox{.}}{2008}]%
        {ebrahimi2008automatic}
\bibfield{author}{\bibinfo{person}{Farideh Ebrahimi}, \bibinfo{person}{Mohammad
  Mikaeili}, \bibinfo{person}{Edson Estrada}, {and} \bibinfo{person}{Homer
  Nazeran}.} \bibinfo{year}{2008}\natexlab{}.
\newblock \showarticletitle{Automatic sleep stage classification based on EEG
  signals by using neural networks and wavelet packet coefficients}. In
  \bibinfo{booktitle}{{\em 2008 30th Annual International Conference of the
  IEEE Engineering in Medicine and Biology Society}}. IEEE,
  \bibinfo{pages}{1151--1154}.
\newblock


\bibitem[\protect\citeauthoryear{Engemann, Raimondo, King, Rohaut, Louppe,
  Faugeras, Annen, Cassol, Gosseries, Fernandez-Slezak, et~al\mbox{.}}{Engemann
  et~al\mbox{.}}{2018}]%
        {engemann2018robust}
\bibfield{author}{\bibinfo{person}{Denis~A Engemann}, \bibinfo{person}{Federico
  Raimondo}, \bibinfo{person}{Jean-R{\'e}mi King}, \bibinfo{person}{Benjamin
  Rohaut}, \bibinfo{person}{Gilles Louppe}, \bibinfo{person}{Fr{\'e}d{\'e}ric
  Faugeras}, \bibinfo{person}{Jitka Annen}, \bibinfo{person}{Helena Cassol},
  \bibinfo{person}{Olivia Gosseries}, \bibinfo{person}{Diego Fernandez-Slezak},
  {and} \bibinfo{person}{others}.} \bibinfo{year}{2018}\natexlab{}.
\newblock \showarticletitle{Robust EEG-based cross-site and cross-protocol
  classification of states of consciousness}.
\newblock \bibinfo{journal}{{\em Brain\/}} \bibinfo{volume}{141},
  \bibinfo{number}{11} (\bibinfo{year}{2018}), \bibinfo{pages}{3179--3192}.
\newblock


\bibitem[\protect\citeauthoryear{Hosseinifard, Moradi, and
  Rostami}{Hosseinifard et~al\mbox{.}}{2013}]%
        {hosseinifard2013classifying}
\bibfield{author}{\bibinfo{person}{Behshad Hosseinifard},
  \bibinfo{person}{Mohammad~Hassan Moradi}, {and} \bibinfo{person}{Reza
  Rostami}.} \bibinfo{year}{2013}\natexlab{}.
\newblock \showarticletitle{Classifying depression patients and normal subjects
  using machine learning techniques and nonlinear features from EEG signal}.
\newblock \bibinfo{journal}{{\em Computer methods and programs in
  biomedicine\/}} \bibinfo{volume}{109}, \bibinfo{number}{3}
  (\bibinfo{year}{2013}), \bibinfo{pages}{339--345}.
\newblock


\bibitem[\protect\citeauthoryear{Ieracitano, Mammone, Hussain, and
  Morabito}{Ieracitano et~al\mbox{.}}{2020}]%
        {ieracitano2020novel}
\bibfield{author}{\bibinfo{person}{Cosimo Ieracitano}, \bibinfo{person}{Nadia
  Mammone}, \bibinfo{person}{Amir Hussain}, {and} \bibinfo{person}{Francesco~C
  Morabito}.} \bibinfo{year}{2020}\natexlab{}.
\newblock \showarticletitle{A novel multi-modal machine learning based approach
  for automatic classification of EEG recordings in dementia}.
\newblock \bibinfo{journal}{{\em Neural Networks\/}}  \bibinfo{volume}{123}
  (\bibinfo{year}{2020}), \bibinfo{pages}{176--190}.
\newblock


\bibitem[\protect\citeauthoryear{Jo and Jeong}{Jo and Jeong}{2020}]%
        {jo2020prediction}
\bibfield{author}{\bibinfo{person}{Sang-Yeong Jo} {and}
  \bibinfo{person}{Jin-Woo Jeong}.} \bibinfo{year}{2020}\natexlab{}.
\newblock \showarticletitle{Prediction of visual memorability with EEG signals:
  A comparative study}.
\newblock \bibinfo{journal}{{\em Sensors\/}} \bibinfo{volume}{20},
  \bibinfo{number}{9} (\bibinfo{year}{2020}), \bibinfo{pages}{2694}.
\newblock


\bibitem[\protect\citeauthoryear{Karis, Fabiani, and Donchin}{Karis
  et~al\mbox{.}}{1984}]%
        {karis1984p300}
\bibfield{author}{\bibinfo{person}{Demetrios Karis}, \bibinfo{person}{Monica
  Fabiani}, {and} \bibinfo{person}{Emanuel Donchin}.}
  \bibinfo{year}{1984}\natexlab{}.
\newblock \showarticletitle{“P300” and memory: Individual differences in
  the von Restorff effect}.
\newblock \bibinfo{journal}{{\em Cognitive Psychology\/}} \bibinfo{volume}{16},
  \bibinfo{number}{2} (\bibinfo{year}{1984}), \bibinfo{pages}{177--216}.
\newblock


\bibitem[\protect\citeauthoryear{Kiziltepe, Constantin, Demarty, Healy, Fosco,
  Garc\'ia Seco~de Herrera, Halder, Ionescu, Matran-Fernandez, Smeaton, and
  Sweeney}{Kiziltepe et~al\mbox{.}}{2021}]%
        {ME2021}
\bibfield{author}{\bibinfo{person}{Rukiye~Savran Kiziltepe},
  \bibinfo{person}{Mihai~Gabriel Constantin},
  \bibinfo{person}{Claire-H\'el\`ene Demarty}, \bibinfo{person}{Graham Healy},
  \bibinfo{person}{Camilo Fosco}, \bibinfo{person}{Alba Garc\'ia Seco~de
  Herrera}, \bibinfo{person}{Sebastian Halder}, \bibinfo{person}{Bogdan
  Ionescu}, \bibinfo{person}{Ana Matran-Fernandez}, \bibinfo{person}{Alan~F.
  Smeaton}, {and} \bibinfo{person}{Lorin Sweeney}.}
  \bibinfo{year}{2021}\natexlab{}.
\newblock \showarticletitle{Overview of The {MediaEval} 2021 Predicting Media
  Memorability Task}. In \bibinfo{booktitle}{{\em Working Notes Proceedings of
  the {MediaEval} 2021 Workshop}}.
\newblock


\bibitem[\protect\citeauthoryear{Klimesch}{Klimesch}{1999}]%
        {klimesch1999eeg}
\bibfield{author}{\bibinfo{person}{Wolfgang Klimesch}.}
  \bibinfo{year}{1999}\natexlab{}.
\newblock \showarticletitle{EEG alpha and theta oscillations reflect cognitive
  and memory performance: a review and analysis}.
\newblock \bibinfo{journal}{{\em Brain research reviews\/}}
  \bibinfo{volume}{29}, \bibinfo{number}{2-3} (\bibinfo{year}{1999}),
  \bibinfo{pages}{169--195}.
\newblock


\bibitem[\protect\citeauthoryear{Lehmann, Koenig, Jelic, Prichep, John,
  Wahlund, Dodge, and Dierks}{Lehmann et~al\mbox{.}}{2007}]%
        {lehmann2007application}
\bibfield{author}{\bibinfo{person}{Christoph Lehmann}, \bibinfo{person}{Thomas
  Koenig}, \bibinfo{person}{Vesna Jelic}, \bibinfo{person}{Leslie Prichep},
  \bibinfo{person}{Roy~E John}, \bibinfo{person}{Lars-Olof Wahlund},
  \bibinfo{person}{Yadolah Dodge}, {and} \bibinfo{person}{Thomas Dierks}.}
  \bibinfo{year}{2007}\natexlab{}.
\newblock \showarticletitle{Application and comparison of classification
  algorithms for recognition of Alzheimer's disease in electrical brain
  activity (EEG)}.
\newblock \bibinfo{journal}{{\em Journal of neuroscience methods\/}}
  \bibinfo{volume}{161}, \bibinfo{number}{2} (\bibinfo{year}{2007}),
  \bibinfo{pages}{342--350}.
\newblock


\bibitem[\protect\citeauthoryear{Liang, Saratchandran, Huang, and
  Sundararajan}{Liang et~al\mbox{.}}{2006}]%
        {liang2006classification}
\bibfield{author}{\bibinfo{person}{Nan-Ying Liang},
  \bibinfo{person}{Paramasivan Saratchandran}, \bibinfo{person}{Guang-Bin
  Huang}, {and} \bibinfo{person}{Narasimhan Sundararajan}.}
  \bibinfo{year}{2006}\natexlab{}.
\newblock \showarticletitle{Classification of mental tasks from EEG signals
  using extreme learning machine}.
\newblock \bibinfo{journal}{{\em International journal of neural systems\/}}
  \bibinfo{volume}{16}, \bibinfo{number}{01} (\bibinfo{year}{2006}),
  \bibinfo{pages}{29--38}.
\newblock


\bibitem[\protect\citeauthoryear{Noh, Herzmann, Curran, and de~Sa}{Noh
  et~al\mbox{.}}{2014}]%
        {noh2014using}
\bibfield{author}{\bibinfo{person}{Eunho Noh}, \bibinfo{person}{Grit Herzmann},
  \bibinfo{person}{Tim Curran}, {and} \bibinfo{person}{Virginia~R de Sa}.}
  \bibinfo{year}{2014}\natexlab{}.
\newblock \showarticletitle{Using single-trial EEG to predict and analyze
  subsequent memory}.
\newblock \bibinfo{journal}{{\em NeuroImage\/}}  \bibinfo{volume}{84}
  (\bibinfo{year}{2014}), \bibinfo{pages}{712--723}.
\newblock


\bibitem[\protect\citeauthoryear{Pedregosa, Varoquaux, Gramfort, Michel,
  Thirion, Grisel, Blondel, Prettenhofer, Weiss, Dubourg,
  et~al\mbox{.}}{Pedregosa et~al\mbox{.}}{2011}]%
        {pedregosa2011scikit}
\bibfield{author}{\bibinfo{person}{Fabian Pedregosa}, \bibinfo{person}{Ga{\"e}l
  Varoquaux}, \bibinfo{person}{Alexandre Gramfort}, \bibinfo{person}{Vincent
  Michel}, \bibinfo{person}{Bertrand Thirion}, \bibinfo{person}{Olivier
  Grisel}, \bibinfo{person}{Mathieu Blondel}, \bibinfo{person}{Peter
  Prettenhofer}, \bibinfo{person}{Ron Weiss}, \bibinfo{person}{Vincent
  Dubourg}, {and} \bibinfo{person}{others}.} \bibinfo{year}{2011}\natexlab{}.
\newblock \showarticletitle{Scikit-learn: Machine learning in Python}.
\newblock \bibinfo{journal}{{\em the Journal of machine Learning research\/}}
  \bibinfo{volume}{12} (\bibinfo{year}{2011}), \bibinfo{pages}{2825--2830}.
\newblock


\bibitem[\protect\citeauthoryear{Rugg and Curran}{Rugg and Curran}{2007}]%
        {rugg2007event}
\bibfield{author}{\bibinfo{person}{Michael~D Rugg} {and} \bibinfo{person}{Tim
  Curran}.} \bibinfo{year}{2007}\natexlab{}.
\newblock \showarticletitle{Event-related potentials and recognition memory}.
\newblock \bibinfo{journal}{{\em Trends in cognitive sciences\/}}
  \bibinfo{volume}{11}, \bibinfo{number}{6} (\bibinfo{year}{2007}),
  \bibinfo{pages}{251--257}.
\newblock


\bibitem[\protect\citeauthoryear{Sanquist, Rohrbaugh, Syndulko, and
  Lindsley}{Sanquist et~al\mbox{.}}{1980}]%
        {sanquist1980electrocortical}
\bibfield{author}{\bibinfo{person}{Thomas~F Sanquist}, \bibinfo{person}{John~W
  Rohrbaugh}, \bibinfo{person}{Karl Syndulko}, {and} \bibinfo{person}{Donald~B
  Lindsley}.} \bibinfo{year}{1980}\natexlab{}.
\newblock \showarticletitle{Electrocortical signs of levels of processing:
  Perceptual analysis and recognition memory}.
\newblock \bibinfo{journal}{{\em Psychophysiology\/}} \bibinfo{volume}{17},
  \bibinfo{number}{6} (\bibinfo{year}{1980}), \bibinfo{pages}{568--576}.
\newblock


\bibitem[\protect\citeauthoryear{Wang, Nie, and Lu}{Wang et~al\mbox{.}}{2014}]%
        {wang2014emotional}
\bibfield{author}{\bibinfo{person}{Xiao-Wei Wang}, \bibinfo{person}{Dan Nie},
  {and} \bibinfo{person}{Bao-Liang Lu}.} \bibinfo{year}{2014}\natexlab{}.
\newblock \showarticletitle{Emotional state classification from EEG data using
  machine learning approach}.
\newblock \bibinfo{journal}{{\em Neurocomputing\/}}  \bibinfo{volume}{129}
  (\bibinfo{year}{2014}), \bibinfo{pages}{94--106}.
\newblock


\end{thebibliography}

\end{document}